\documentclass[prb,twocolumn,showpacs,preprintnumbers,amsmath,amssymb,showkeys]{revtex4}
\usepackage{graphicx}
\usepackage{psfrag}
\usepackage{epsfig}

\begin{document}
\title{Fluctuation induced hopping and spin polaron transport}
\author{L.~G.~L.~Wegener} 
\author{P.~B.~Littlewood}
\affiliation{TCM, Cavendish Laboratory, Cambridge University,\\ 
             Madingley Road, Cambridge, CB3 0HE, United Kingdom}
\date{\today}

\begin{abstract} 
We study the motion of free magnetic polarons in a paramagnetic background 
of fluctuating local moments.  The polaron can tunnel only to nearby 
regions of local moments when these fluctuate into alignment.  We propose 
this fluctuation induced hopping as a new transport mechanism for the spin 
polaron.  We calculate the diffusion constant for  fluctuation induced 
hopping from the rate at which local moments fluctuate into alignment. The 
electrical resistivity is then obtained via the Einstein relation.  We 
suggest that the proposed transport mechanism is relevant in the high 
temperature phase of the Mn pyrochlore colossal magneto resistance 
compounds and EuB$_6$.  
\end{abstract} 

\maketitle

Recently Free Magnetic Polarons (FMP) have received renewed attention.  
They were proposed to explain the Colossal Magnetoresistance (CMR) in the 
Manganese Pyrochlore 
compounds\cite{MajumdarLittlewood1998a,MajumdarLittlewood1998b,CalderonLittlewood1999,martinez1999,ShimikawaKuboManako1996,RamirezSubramanian1997} 
and they have been studied in the context of the double exchange model and 
the Manganese Perovskite CMR 
compounds\cite{KaganKhomskiiMostovoy2000,MoreoYunokiDagotto1999,AllodiGenziGuidi1998}.  
Moreover, Raman scattering data has suggested\cite{SnowCooper2000,NyhusYoon1997} 
that they exist in EuB$_6$.  Previous theoretical studies have focussed 
on the static properties of the FMP. Here we focus on the dynamic aspect, 
propose a new transport mechanism for a FMP and calculate the resulting 
resistivity.  

A magnetic polaron is a composite object consisting of a localised charge 
carrier and the alignment it induces in a background of local moments.  
Localisation can occur for two different reasons: the carrier can be 
trapped by an impurity atom and then induce a magnetisation in the region 
where it is localised. The resulting particle is called a ``Bound Magnetic 
Polaron'' (BMP). It is well documented experimentally, for example in 
dilute magnetic semiconductors like Cd$_{1-x}$Mn$_x$Se 
\cite{IsaacsHeiman1988}, and in rare earth 
chalcogenides\cite{vonMolnarandMethfessel1967}.  It has been studied in 
depth theoretically\cite{Wolff1988}. A BMP is not free to roam through the 
sample since it is bound to its impurity. Only activated transport is 
possible: when the BMP is ``ionised'' the carrier is free to move until 
it is trapped by the next impurity. 

However, for large enough coupling to the local moments the carrier can 
self-trap without the need for an 
impurity\cite{KasuyaYanaseTakeda1970,Nagaev1971,Guillaume1993}, forming a 
Free Magnetic Polaron (FMP). Due to the coupling the carrier acts as a 
magnetic field on the local moments. The strength of this field varies in 
space as the probability density of the carrier: the more localised the 
carrier the stronger the field and the larger the energy gain resulting from 
aligning the local moments.  The region of aligned moments 
therefore acts as a potential well that localises the electron and a FMP is 
formed.  The balance between the gain in magnetic energy from induced 
alignment and the loss in kinetic energy because of localisation determines 
the polaron size. The existence of an FMP has not been confirmed 
experimentally, but it has been suggested in the 
Mn-pyrochlores\cite{martinez1999,ShimikawaKuboManako1996,RamirezSubramanian1997} 
and in EuB$_6$ as well\cite{SnowCooper2000}. 

\begin{figure}
\psfrag{En}{$E$}
\psfrag{Ep}{$E_\mathrm{p}$}
\psfrag{gE}{$g(E)$}
\psfrag{tA}{$\tau_{\mathrm{A}}$}
\psfrag{tX}{$\tau_{\mathrm{X}}$}
\psfrag{tS}{$\tau_{\mathrm{S}}$}
\psfrag{tH}{$\tau_{\mathrm{hop}}$}
\psfrag{lt}{$\log(t)$}
\psfrag{lc}{Level crossing}
\psfrag{ah}{and hop}
\includegraphics[width=3.3in]{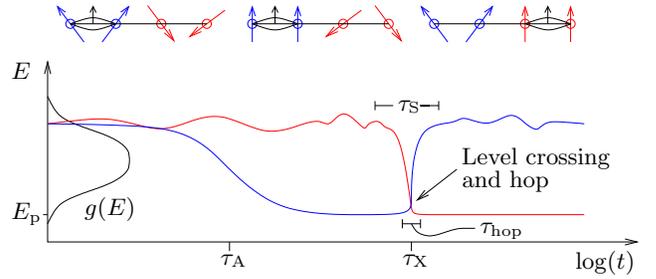}
\caption{\label{fig:Timeline}\textbf{Time evolution of electron state.} 
Black curved lines and arrows in the upper part represent the electron 
density and spin.  A carrier is put on the ``blue'' level.  A hop occurs 
when the ``red'' and ``blue'' levels cross.The density of states $g(E)$ 
is shown on the LHS of the lower part.}
\end{figure}

The mechanism of transport by FMP is in doubt. The conventional view is 
that transport is necessarily activated, as for a BMP. Here we present an 
alternative viewpoint: we propose that, unlike the BMP, the FMP can move 
between nearby sites without thermal assistance.  We consider a 
fluctuating, paramagnetic background of local moments.  A neighbouring 
region of local moments can fluctuate into the same alignment as the 
polaronic moments.  At that moment, the carrier can tunnel to the newly 
aligned region without needing to overcome an energy barrier. The tunneling 
process is fast compared to the spin fluctuations. After the tunneling 
process the carrier and the alignment have moved so that the complete FMP 
has hopped to the new location.  The entire time evolution of the polaron 
formation and hopping process is illustrated in Figure \ref{fig:Timeline}.  
We call this  transport mechanism ``Fluctuation Induced Hopping'' (FIH).  
It does not involve an activated process. We calculate the 
resistivity FIH gives rise to.

In the next section we present a model Hamiltonian that provides the 
frame of reference for our work. We describe the static properties of the FMP
and the band states in Sec. \ref{sec:small}, postponing a justification 
until Appendices \ref{varcalcpol} and \ref{varcalcband}.  In 
Sec.~\ref{sec:FIH} we calculate the electrical resistivity for polaron 
hopping.  We determine the rate at which nearby regions of local moments 
accidentally align themselves due to fluctuations.  Since the FMP
tunnels to these regions, this rate determines the diffusion constant and 
hence the electrical resistivity.  

\section{Model Hamiltonian}
We consider a low density electron gas that is coupled ferromagnetically to 
a background of local moments. The local moments are themselves coupled ferromagnetically.
The following Hamiltonian describes this 
system\cite{MajumdarLittlewood1998a,MajumdarLittlewood1998b}:
\begin{equation}
H= t\!\!\!\sum_{\langle i,j\rangle\sigma}\!\!\!c_{i\sigma}^\dagger c_{j\sigma} 
   - J'\sum_{i}\vec{\sigma}_i \cdot \vec{S}_i 
   -J\sum_{\langle i,j\rangle}\vec{S}_i \cdot \vec{S}_j .
\label{petersH}
\end{equation}
Here $i$ denotes the lattice site, $c_{i\sigma}^\dagger$ creates a 
conduction electron, $\vec{S}_i$ is the local moment on site $i$ and  
$\vec{\sigma}_i$ is the conduction electron's spin.  $\langle i,j \rangle$ 
denotes a summation over nearest neighbours.  The first term of the 
Hamiltonian in Eqn.~\ref{petersH} is the kinetic energy of the carriers, 
the second term couples the carriers to the local moments on which they 
reside.  The third term couples the local moments ferromagnetically.  This 
term can be due to, for example, superexchange.  We have an $s$-$d$ 
Hamiltonian with an additional Heisenberg term.  We consider the strong 
coupling regime in which
\begin{equation}
J'\gtrsim t \sim 10J \sim 0.1\, e{\mathrm{V}}.
\label{relations}
\end{equation}
In our calculations we use the values $J'=5t$ and $J=0.01$eV which are in 
agreement with experimental values in the relevant 
materials\cite{MajumdarLittlewood1998a,SnowCooper2000}.  It should be noted 
that the magnetic transition is not driven by the $s$-$d$ part of the 
Hamiltonian, but only by the superexchange because of the low carrier density.

\section{Polaron and Band State} 
\label{sec:small}
Here we present the wave function we use to study the transport 
properties of the FMP.  Our variational calculation in Appendix 
\ref{varcalcpol} shows that the FMP is small in the strong coupling 
regime: the carrier occupies approximately two lattice sites (see Figure 
\ref{fig:pgraph}). We therefore use the following wave function to describe 
the polaron:
\begin{equation}
\left|P\right\rangle =
 \frac{1}{\sqrt{2}}( c_0^\dagger + c_1^\dagger)\left|0\right\rangle\otimes\left|\uparrow\right\rangle_{\vec{m}}\otimes\left|\vec{m}\right\rangle,
\label{smallpolaronwavefunction}
\end{equation}
where $\vec{S}_i$ are the polaronic local moments and 
$\vec{m} = \vec{S}_0 + \vec{S}_1$ is the magnetisation of the FMP.  It 
describes a carrier localised on two lattice sites, with its spin quantised 
and pointing ``up'' along the direction of $\vec{m}$ to minimise the $s$-$d$ 
energy.  The $s$-$d$ term in the Hamiltonian therefore reduces to 
$-J'\sigma m$, which is lower for more aligned local moments.  This means 
that the carrier introduces an additional coupling between the polaronic 
moments.  The value of the effective coupling constant can be obtained by 
expanding the expression for $m$ up to first order in 
$\vec{S}_0\cdot\vec{S}_1 - S^2$ for nearly aligned polaronic moments.  We 
obtain
\begin{equation}
J_{\mathrm{eff}} = \frac{J'}{\sqrt{1+8S(2S+1)}} + 2J.
\label{Jeff}
\end{equation}
Since $T\ll J'$ the moments are nearly aligned, so the polaron energy is
\begin{equation}
E_{\mathrm{p}}= -|t| - (J'\sqrt{2S(2S+1)}/2 +2JS^2).
\label{2siteenergy}
\end{equation}

In addition to the polaron state with its induced magnetisation there are 
many more possible states for the carrier in which it does not align any 
local moments.  These are states in the narrowed band described in Ref.~\onlinecite{KoganAuslender1988}.  Since the background fluctuates, these 
``band states'' persist at a given location only for a time comparable to 
the timescale of these fluctuations, which we denote $\tau_\mathrm{S}$.  
Nevertheless, the FMP would be unstable if a significant number of lower 
energy band states existed, since the carrier could then tunnel to them and 
gain energy.  

To check whether the FMP is stable we need to estimate the position of the 
band edge. If the polaron level lies below the band edge, band states with 
a lower energy are exceedingly rare, and can be neglected. If on the other 
hand the band edge lies below the energy of the FMP, the latter is 
unstable. We determine the position of the band edge as the lowest energy 
of a typical band state.  We use a variational approach to calculate this 
energy (details are given in Appendix \ref{varcalcband}).  The size of the 
band state is determined by the balance of the kinetic energy cost of 
localisation and the gain from the $s$-$d$ term.  The latter is very small: 
the carrier aligns its spin with the total magnetisation of the region it 
is localised in. This magnetisation is due to statistical fluctuations in 
the paramagnetic regime, and consequently very small.  We therefore expect 
the kinetic term to dominate and the band state to be very large.  This is 
confirmed by our calculation.  We show that typical band states are 
extended over a region of roughly 10$^6$ lattice sites. Their total energy 
is very close to $-6t$ (See Equation (\ref{largebandstateenergy})).  
Therefore the polaron level lies below the band edge in the regime we 
consider ($J'\gtrsim t$),  and the FMP is stable. This concludes our 
discussion of the static properties of the FMP and the band state. In 
the next section we determine their dynamic properties.

\section{Fluctuation Induced Hopping} 
\label{sec:FIH} 
In this section we consider the FIH mechanism in detail and calculate the 
hopping rate of the polaron and the resulting electrical resistivity. Let us 
examine the time evolution of a single carrier that is injected onto a 
lattice site in an empty system.  The local moments in its vicinity cannot 
respond immediately to the carrier's presence, but react on a timescale 
$\tau_\mathrm{A}$. On  timescales smaller than $\tau_\mathrm{A}$, the 
background appears static.  A completely static background would cause 
Anderson localisation\cite{KoganAuslender1988} and trap the carrier in a 
state in the tail of the band below the mobility edge.  Although localised, 
the carrier has not yet induced any alignment between local moments in its 
vicinity; it is in one of the band states described above.  Only when the 
moments have aligned themselves is the energy of the state greatly reduced.  
This means that there is a large energy barrier that prevents the carrier 
from making thermal hops out of the region of alignment.  However, the 
energy of band states fluctuates because the local moments fluctuate.  It 
can therefore cross the polaron level.  At such a crossing the electron can 
tunnel to this level, since there is no more energy barrier to overcome.  
We call this tunnelling process Fluctuation Induced Hopping (FIH).  After 
the electron tunnelled, the entire FMP has moved: carrier and alignment 
have jumped to another site.  We summarise the entire time evolution in 
Fig.~\ref{fig:Timeline}.

The occurrences of hops are uncorrelated in time and space since the 
background of local moments is paramagnetic.  FIH is therefore a Markoff 
process and the FMP executes a random walk.  In an ensemble of 
realisations of the polaron and the background, polarons in different 
realisations follow different paths.  The Probability Density Function 
(PDF) of the polaron, defined as the fraction of realisations in an 
ensemble that have the FMP at a specified time at a specified location, 
obeys the diffusion equation.\cite{Chandrasekhar1943} The diffusion 
constant in this equation characterises the polaron transport in the long 
time limit.

For a random walk in three dimensions that consists of hops of $l$ lattice 
constants, occurring with a frequency $\omega_l$,  the diffusion constant 
is:\cite{Chandrasekhar1949} 
\begin{equation}
D=\frac{1}{6}\sum_{l=1}^{\infty} (l a)^2 \omega_l,
\label{fundamentaldiffconst}
\end{equation} 
where $a$ is the lattice constant.  The resistivity from polaron transport 
is then obtained from the diffusion constant by means of the Einstein formula:
\begin{equation}
\rho=(n e\mu)^{-1}=\frac{k_B T}{n e^2 D},
\label{resistivityanddiffconst}
\end{equation}
where $\mu$ is the mobility, and $n$ the number density of polarons.

\subsection{Rate of level crossings} \label{subsec:levelcross} We calculate 
the rate at which band state levels cross the polaron level.  Such a 
crossing occurs when the band state energy fluctuates so much that it lies 
at or below the polaron level.  The crossing rate depends on the size of 
the band state: the energy gap between the FMP and band states depends 
on their size.  In addition to this,  we will see that the energy of large 
band states fluctuates less.  First we calculate the crossing rate for 
small band states.

The energy difference between the polaron and a small band state is only 
due to the difference in exchange energy since their kinetic energies are 
identical.  The time at which the RMS deviation of the band state energy 
becomes as large as the energy gap between the two levels therefore 
determines the crossing rate.
\begin{eqnarray}
(\langle \vec{S}_0 \cdot\vec{S}_1 \rangle_{J_{\mathrm{eff}}} - \langle \vec{S}_0 \cdot\vec{S}_1 \rangle_{J})^2
&=&\nonumber\\
\langle [\vec{S}_0(\tau_{\mathrm{X}}) \cdot\vec{S}_1(\tau_{\mathrm{X}}) - \vec{S}_0(0) \cdot\vec{S}_1 (0) \rangle_{J}]^2 \rangle,&&
\end{eqnarray}
where angular brackets denote ensemble averaging.  We neglect the variance 
of the polaronic exchange energy since it is very small due to the large 
effective coupling constant of the polaronic moments (See Appendix 
\ref{sec:small}).  We also neglect the exchange energy of the band state 
compared to the polaron exchange energy:
\begin{eqnarray}
\langle[\vec{S}_0(\tau_{X})\cdot \vec{S}_1(\tau_{X})]
[\vec{S}_0 \cdot\vec{S}_1]\rangle_{J} 
=\nonumber \\ 
\langle[\vec{S}_0\cdot \vec{S}_1]^2\rangle_{J} -\frac{1}{2}\langle\vec{S}_0\cdot \vec{S}_1\rangle_{J_{\mathrm{eff}}}^2.
\label{equationfortaux2}
\end{eqnarray}
For simplicity we use an interpolation between $t=0$ and $t=\infty$ instead 
of the exact analytical form of the four-point correlator in 
Eqn.~\ref{equationfortaux2}.  At time $t=0$  the correlator reduces to 
$\langle (\vec{S}_0 \cdot \vec{S}_1)^2\rangle$.  A pair of spins at $t=0$ 
is completely uncorrelated with itself at $t=\infty$, so that in this limit 
the correlator reduces to $\langle\vec{S}_0\cdot\vec{S}_1\rangle^2$.  We 
can therefore interpolate as follows:
\begin{eqnarray}
\langle[\vec{S}_0(t)\cdot\vec{S}_1(t)]
            [\vec{S}_0(0) \cdot \vec{S}_1(0)] \rangle
=\nonumber \\ 
\langle (\vec{S}_0\cdot\vec{S}_1)^2\rangle 
  - f(\frac{t}{\tau_{\mathrm{S}}})\left[
              \langle(\vec{S}_0\cdot\vec{S}_1)^2\rangle
             -\langle \vec{S}_0\cdot\vec{S}_1  \rangle^2
        \right],
\label{def of f}
\end{eqnarray}
where $f(x)$ varies smoothly from 0 at $x=0$ to 1 at $x=\infty$.  Moreover, 
we have written $f(t/\tau_{\mathrm{S}})$ since the four-point correlator 
varies on the same timescale as the fluctuations of the background. 
 This allows us to rewrite Eqn.~\ref{equationfortaux2} as:
\begin{equation}
f(\frac{t}{\tau_{\mathrm{S}}})=\frac{\frac{1}{2}\langle \vec{S}_0\cdot
\vec{S}_1\rangle_{J_{\mathrm{eff}}}^2}{\langle(\vec{S}_0\cdot\vec{S}_1)^2\rangle_J 
- \langle \vec{S}_0\cdot\vec{S}_1 \rangle_J^2}.
\end{equation}
Since $f(x)$ is a smooth function varying between $0$ and $1$ that changes 
mostly near $x=1$,  $f(1) \approx 1/2$ and $f'(1) \approx 1$.  We expand 
$f(t/\tau_{\mathrm{S}})$ up to first order about $t=\tau_{\mathrm{S}}$, 
which yields 
\begin{equation}
\frac{\tau_{\mathrm{X}}}{\tau_{\mathrm{S}}} \propto 
\frac{\langle \vec{S}_0\cdot\vec{S}_1\rangle_{J_{\mathrm{eff}}}^2}
{\langle(\vec{S}_0\cdot\vec{S}_1)^2\rangle_J 
 - \langle \vec{S}_0\cdot\vec{S}_1 \rangle_J^2}.
\label{tauX1}
\end{equation}
This means that crossings occur on the time scale of the fluctuations of 
the Heisenberg magnet weighted by the different alignments of the polaron 
and the background spins.  Since $k_B T \ll J'$, the numerator of 
Eqn.~\ref{tauX1} reduces to $(S(S+1))^2$. 

For $k_\mathrm{B}T\gtrsim\mathcal{O}(J)$ the two terms in the denominator 
of Eqn.~\ref{tauX1} have very similar temperature dependences, so that 
\begin{equation}
\tau_{X} = A\, \tau_{\mathrm{S}} 
           \frac{(S(S+1)^2}
                {\langle\vec{S}_0 \cdot \vec{S}_1\rangle_J^2},
\label{tauX2}
\end{equation}
where $A \sim {\mathcal{O}}(1)$. 
We use the expression given in Ref. \onlinecite{Lovesey1986} in the denominator 
and  $\tau_{\mathrm{S}} = \hbar \sqrt{\beta / J}$.\cite{Hubbard71}
to obtain 
\begin{equation}
\tau_{X} =A\frac{\hbar}{J} (\beta J)^{-\frac{3}{2}}.
\label{tauX3a}
\end{equation}
$\tau_{X}$ increases with temperature.  With increasing temperature level 
crossings become more rare.  The decrease in the timescale of the spin 
fluctuations is more than offset by the increase of the average 
misalignment of the local moments.

For $k_\mathrm{B}T\gg J$ the temperature dependences differ: 
$\langle(\vec{S}_0\cdot\vec{S}_1)^2\rangle$ tends to $S^2/3$ since the 
spins are completely uncorrelated, whereas 
$\langle\vec{S}_0 \cdot \vec{S}_1\rangle_J $ vanishes as $1/T$.  Moreover 
the timescale of the fluctuations is different:\cite{BlumeHubbard1970} 
$\tau_\mathrm{S}=\hbar/\sqrt{S(S+1)}J$. Hence $\tau_\mathrm{X}$ tends to a 
constant in this limit:
\begin{equation}
\label{tauX3b}
\tau_\mathrm{X} = \frac{3\hbar}{J\sqrt{S(S+1)}}.
\end{equation}
The reason is that the local moments are completely disordered in this 
regime; an increase in temperature does not cause an increase in disorder.
The crossover between the two regimes occurs at a temperature of about $JS(S+1)/k_\mathrm{B}$.

Equations (\ref{tauX3a}) and (\ref{tauX3b}) constitute the principal 
result of this section.  However, before proceeding to the estimate of the 
diffusion constant we should check that fluctuations of other band states - 
in particular those involving rearrangements of many spins - do not change 
our conditions.  We calculate the crossing rate of large band states with 
the polaron level in the same way as before (see Appendix \ref{sec:large 
band states} for details).  For a level crossing with a large band state, 
many local moments need to fluctuate into alignment simultaneously. This is 
a very unlikely event, and one expects the crossing rate to be accordingly 
small. This expectation is borne out by our calculation. In Appendix 
\ref{sec:large band states} we show that level crossings with large band 
states can be neglected safely.

\subsection{Diffusion Constant} \label{sec:diffusion const} 
\begin{figure}
\psfrag{7}{\raisebox{-1mm}{$T_\mathrm{c}/J$}}
\includegraphics[width=3.3in]{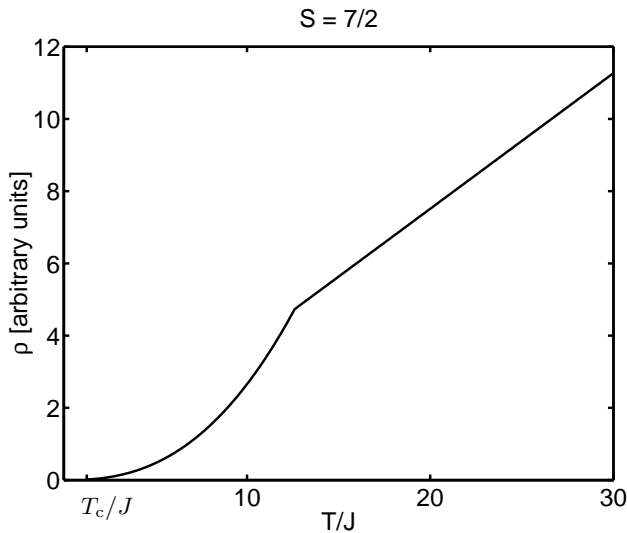}
\caption{\label{fig:rhovst}\textbf{Temperature dependence of the resistivity.}}
\end{figure}

When a crossing 
occurs it is possible, but not necessary, for the FMP to hop.  The 
probability, ${\mathcal{P}}_l$, of a hop of length $l$ at a level crossing 
depends on the overlap between polaron and band state wave functions and on 
the rate at which the levels cross.  The frequency at which hops of length 
$l$ occur, $\omega_l$, is then 
\begin{equation}
\omega_l \propto  l^2 \tau_{X}^{-1} {\mathcal{P}}_l,
\label{expression for omega n}
\end{equation}
since the number of small band states a distance $l$ away from the FMP
is roughly proportional to $l^2$.  The tunnelling probability from the 
polaron state, $\left|P\right\rangle$, to the band state, 
$\left|B\right\rangle$ with energies $E_{\mathrm{P}}$ and $E_{\mathrm{B}}$ 
is given by the Landau-Zener formula \cite{Zener1932,Landau1932}
\begin{equation}
{\mathcal{P}}_{P\rightarrow B} = 1 - \exp\left[-\frac{2\pi}{\hbar}
\frac{\left|\left\langle P\right|H\left|B\right\rangle\right|^2}
     {\left|\frac{\partial}{\partial t} (E_{\mathrm{P}}-E_{\mathrm{B}})\right|}\right],
\label{zenerstransitionprob}
\end{equation}
where $H$ is the Hamiltonian for the particle.
Here we have $\left|\frac{\partial}{\partial t} (E_{\mathrm{P}}-E_{\mathrm{B}})\right|\sim J'S/\tau_{\mathrm{S}}$,
since the energy difference is due to the initially unaligned local moments 
in the band state.  The time in which this difference between the levels 
disappears is $\tau_{\mathrm{S}}$.  The spatial extent of the wave 
functions of the polaron and the small band state limits the hopping range 
to one lattice constant.  The overlap between neighbouring small polaron 
states given is by:
\begin{equation}
\langle A | H | B\rangle  =  
\frac{1}{2}\langle 0 | (c_0 + c_1)H(c^\dagger_1 + c^\dagger_2)|0\rangle
=-t -\frac{J'}{2},
\end{equation}
where the sites 0 and 1 are nearest neighbours and 1 and 2 are nearest 
neighbours.  Therefore the hopping probability at a level crossing between 
two neighbouring energy levels is given by:
\begin{eqnarray}
P_{A\rightarrow B} &=& 1 - 
\exp[-2\pi\frac{(t+J'\sqrt{2S(2S+1)}/2)^2}{J J'}\sqrt{\frac{\beta J}{S}}]\nonumber\\
&\approx& 1,
\end{eqnarray}
since $1\ll J'/J$.  This means also that the probability is largely 
independent of temperature.  We will therefore take the hopping probability 
at a level crossing between two neighbouring levels to be 1.  The diffusion 
constant  and resistivity for FIH are therefore 
\begin{equation}
\begin{array}{rcl@{\quad\mathrm{for}\quad}rcl}
D &\propto& a^2 \frac{J}{\hbar}(\beta J)^{\frac{3}{2}}       & k_\mathrm{B}T&\gtrsim& J  \\
D &\propto& \mathrm{const.}                                  & k_\mathrm{B}T&\gg& J  \\
\rho &\propto& \frac{\hbar}{n e^2a^2}\,(k_B T/J)^\frac{5}{2}.& k_\mathrm{B}T&\gtrsim& J  \\
\rho &\propto& \frac{\hbar}{n e^2a^2}\, k_B T/J              & k_\mathrm{B}T&\gg& J  \\
\end{array}
\label{Diffusion constant result 1}
\end{equation}
This is our main result. We plot the resistivity versus temperature in 
Figure \ref{fig:rhovst} interpolating between the high and the extremely 
high temperature regime. Firstly, the FMP can only hop to neighbouring sites 
when a favourable statistical fluctuation aligns the local moments.  These 
fluctuations are statistically likely in the sense that they can be 
estimated from the RMS deviation of the spin fluctuations.  Occurrences of 
alignment far from the polaron do not lead to hopping.  Secondly, while 
small polarons typically hop by this process, large ones cannot. The 
required spin fluctuation into the correct configuration is statistically 
very rare.  Thirdly, the timescale of the spin fluctuation is slow enough 
that the FMP hops with probability 1 once the requisite configuration is 
obtained.  Fourthly, the diffusion constant decreases and the resistivity 
increases as a function of temperature.  This reflects the increasing time 
intervals between the level crossings for higher temperatures and the 
relation between the resistivity and the diffusion constant.

The resistivity we obtained is ``metallic'': it increases with temperature, 
even though we are not considering a metallic system at all.  It is 
interesting to compare our result to the resistivity of a very dirty metal, 
where $k_\mathrm{F} \lambda_\mathrm{f} \approx 1$, $\lambda_\mathrm{f}$ 
being the mean free path for the carriers.  In such a material the Drude 
formula for the resistivity yields:
\begin{equation}
\rho = \frac{\hbar}{n e^2 \lambda_\mathrm{f}^2}.
\label{dirtymetalrho}
\end{equation}
It is clear that despite its temperature dependence the polaron hopping 
resistivity is far too large to be confused with scattering of metallic 
carriers.  The mean free path in the dirty metal would need to approach one 
lattice constant for the resistivities to be comparable.  At such short 
mean free paths the metallic picture of delocalised carriers breaks down.

\section{Conclusions}
\label{sec:comp}
We have proposed a new transport mechanism for FMP in a fluctuating 
disordered background of local moments. In our theory the FMP hops at 
the occurrence of a favourable fluctuation. The transport mechanism is 
therefore not activated, but gives rise to a ``metallic'' resistivity.
Experimentally our theory can be checked by a measurement of the temperature 
dependent resistivity. Such measurements have been performed in 
two systems in which the presence of a FMP has been suggested: the 
Mn-pyrochlores and EuB$_6$.  

In the 
Mn-pyrochlores\cite{RamirezSubramanian1997,CheongHwangBatloggRupp1996,AlonsoMartinezetal1999} 
the resistivity decreases with increasing temperature above $T_\mathrm{c}$.  
This is not in accord with our predictions. There are several reasons for 
this.  There are different types of disorder that affect the resistivity, 
but were not taken into account in our work.  In the In-doped 
compounds \cite{CheongHwangBatloggRupp1996}, there is a miscibility gap for 
dopings of $0.5\le x\le 1.5$.  In this regime the bulk consists of two 
types of grains, each of a distinct phase with a different lattice 
parameter.  Transport is dominated by processes associated with the grain 
boundaries.  Phase separation is suspected to occur in the Sc-doped 
materials as well.  A different type of disorder occurs in Bi-doped 
compounds: a Bi ion introduces a strong-scattering 6$s$ 
vacancy on the Tl-sublattice.  This could bind the polaron, making its 
hopping activated.  Both the scattering centres and the grain boundaries 
make the predicted resistivity difficult to observe.  

EuB$_6$ is a much cleaner system, in which magnetic polarons have been 
observed by spin flip Raman scattering 
experiments.\cite{SnowCooper2000,NyhusYoon1997} There is good qualitative 
agreement with experiment in the high temperature paramagnetic regime. The 
resistivity increases rapidly with temperature up to about 150K and then 
more slowly; at temperatures above 200K the resistivity increases even more 
slowly.\cite{CooleyAronson1997,SullowPrasad1998}  This agrees qualitatively 
with the crossover we predict. The crossover temperature we predict, 225K 
is only a little bit too large. The slowdown above 200K is probably due to 
other scattering mechanisms for the local moments, such as spin-orbit 
coupling and scattering by carriers.  There is no quantitative agreement as 
no $T^{5/2}$ law is observed.  This is probably due to material specific 
complications that our theory does not take into account: EuB$_6$ exhibits 
two distinct magnetic transitions between phases with different magnetic 
anisotropies.\cite{SullowPrasad1998} 

There is good agreement as well with the 2-dimensional Monte Carlo 
simulation\cite{CalderonLittlewood1999}.  The simulation and our work agree 
qualitatively on the static characteristics of the polaron, such as the 
temperature and $J'$ dependence of its size and the core magnetisation.  We 
also agree on the temperature window in which the FMP exists.  We do not 
expect more than qualitative agreement given the different parameter 
regimes that were explored: the simulation uses a much weaker superexchange 
coupling.  The discrepancy between the results for the binding energy in 
the critical regime can be understood.  The simulations show a decrease in 
magnitude with temperature and we predict an increase.  This is due to the 
breakdown of our high temperature approximation.  This also explains why 
the simulation observes a larger polaron: we neglect the  correlations 
between local moments except those induced by the presence of the carrier, 
whereas the simulation takes all correlations into account.

The results for the dynamic characteristics of the FMP are different.  
The simulation shows a diffusion constant that is nearly independent of 
temperature.  This result does not take into account the decrease of the 
timescale of the fluctuations: the diffusion constant is calculated as the 
average hopping per unit Monte Carlo time, defined as the number of 
rediagonalisations.  To correct for this  we divide the numerical diffusion 
constant by $\tau_{\mathrm{S}}$.  This correction increases the discrepancy 
since the diffusion constant now decreases as a function of temperature.
The discrepancy between the corrected Monte Carlo diffusion constant and 
our result indicates that the simulation looks at a slightly different form 
of polaron transport.  In the simulation the polarons are larger than in 
our work.  Larger polarons can move if a shell of neighbouring local 
moments aligns themselves.  This motion ``through accretion'' was not taken 
into account in our result, which was derived for a small FMP.
Our theory and the Monte Carlo simulation do therefore agree qualitatively 
where agreement is expected.

\begin{acknowledgments} We would like to thank Professor Khmelnitskii, Dr.  
B.~D. Simons, Dr.~M.~C\^{o}t\'{e}, Dr.~I.~Smolyarenko, M. Calder\'{o}n, 
M.~L.~Povinelli, P.~Eastham,  and V.~Tripathi for fruitful discussions.  
This work was supported in part by EPSRC and Trinity College Cambridge.  
\end{acknowledgments}

\appendix 
\section{Small spin polarons} 
\label{varcalcpol} 
We determine the size and energy of the FMP in strong coupling regime by a 
variational calculation.  We choose a trial wave function for the carrier 
and obtain the electron density at the local moments, which acts as an 
external field on the moments.  The resulting alignment  and decrease in 
energy is calculated using Curie-Weiss theory.  The expectation value of 
the energy of the trial wave function is then minimised with respect to the 
size of the wave function. 

A Gaussian-like function is used for the electronic part of the trial wave 
function for the FMP; in the notation of Eqn. \ref{petersH}: 
\begin{equation}
\left|P\right\rangle =
 \frac{1}{\sqrt{\mathcal{N}}}
\sum_{\vec{r}_i} e^{-(r_i /\lambda)^2}c_{\vec{r}_i}^\dagger \left|0\right\rangle\otimes\left|\uparrow\right\rangle_{\vec{m}}\otimes\left|\vec{m}\right\rangle,
\label{polaronNsitewavefunction}
\end{equation}
where the vectors $\vec{r}_i$ are  the positions of the lattice sites, 
measured from the centre of the polaron and $ \mathcal{N}$ ensures proper 
normalisation.  $\left|\uparrow\right\rangle_{\vec{m}}$ denotes the 
electrons spin, which is quantised and pointing ``up'' along the direction 
of the average magnetisation induced by the carriers presence.  
$\left|\vec{m}\right\rangle$ is the state vector of the polaronic local 
moments.  The wings of this wave function take into account that a trapped 
electron can nearly always make short excursions to a neighbouring 
non-polaronic local moment.  This is possible since it has nearly always a 
spin component parallel to this moment. These excursions diminish the 
polaron energy insofar as they reduce the magnetisation of the core through 
a reduced electron density. 

\begin{figure}[t]
\begin{center}
\epsfig{file=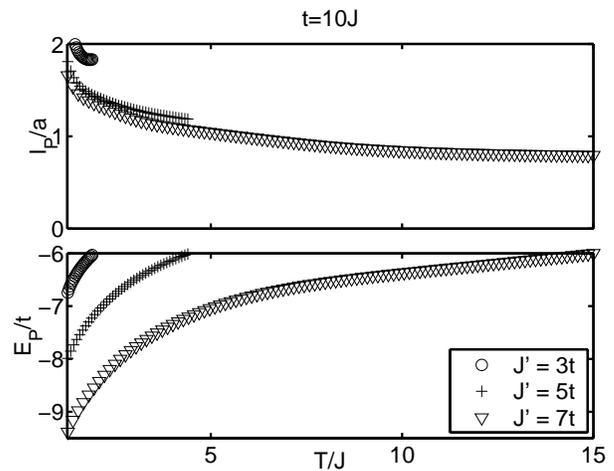,width=8cm}
\end{center}
\caption[Polaron Size end Energy 
vs.~Temperature]{\label{fig:pgraph}\textbf{Polaron Size vs.~Temperature.} 
The curves stop when the polaron energy reaches the edge of the band 
(estimated in Sec. \ref{varcalcband}).} \end{figure}

The magnetisation of the background resulting from the presence of the 
carrier is obtained from Curie-Weiss theory: 
\begin{equation}
m(\vec{r}) = \frac{3}{2} 
\mathrm{B}_{\frac{3}{2}}
[\beta(Jm + \frac{J'}{2}\rho_{\vec{r}_i})],
\label{magnetisation}
\end{equation}
where $\mathrm{B}$ is the Brillouin function and $\rho_{\vec{r}_i}$ is the 
electron density at site $\vec{r}_i$.  Curie-Weiss theory neglects spatial 
correlations between the local moments that are not due to the effective 
field of the carrier and the induced magnetisation itself.  These spatial 
correlations are small in a paramagnet, since the coupling constant of the 
non-polaronic moments is much smaller than the effective field in the core 
of the FMP ($J\ll J'$).  Curie-Weiss theory is therefore accurate in 
the core of the polaron.  In the critical regime non-polaronic correlations 
become important. Our theory is not valid there.

The expectation value of the energy of the trial wave function is minimised 
numerically with respect to the size, $\lambda$, of the polaron.  The 
results are shown below in Fig.~\ref{fig:pgraph}.  We see that the FMP is 
small: its size is of the order of a lattice constant over a wide range of 
temperatures for realistic values of the parameters.  The calculated 
magnetisation is always close to saturation in the core of the polaron, but 
decreases slightly as a function of temperature.  For higher temperatures 
the FMP shrinks and its energy increases.  This is because at constant size 
the energy gain from the $s$-$d$ term decreases as the temperature is 
raised.  To compensate the polaron shrinks, so that the effective 
field due to the carrier increases, which results in a magnetic 
energy gain.  Hence, at higher temperatures, the point of minimum energy is 
shifted towards smaller sizes.  This also means that a smaller $s$-$d$ 
coupling increases the size.  The increase of the polaron energy with 
temperature can be understood as follows:  the kinetic energy is 
independent of temperature and a decreasing function of size.  The magnetic 
energy curve shifts up as temperature is raised and is an increasing 
function of size.  Therefore, the minimum value of the total energy 
increases with temperature.  These results are in close 
agreement with Ref. \onlinecite{MajumdarLittlewood1998a} which 
refers specifically to the pyrochlores.

Our calculation is well-behaved at the ferromagnetic transition, even 
though this temperature lies outside its range of validity.  As 
$T_{\mathrm{c}}$ is approached from above, the polaron size tends to a 
value of approximately $-9.4 t$ at $T_{\mathrm{c}}$.  Neither the present 
calculation, nor Ref.~\onlinecite{MajumdarLittlewood1998a} takes the 
correlations of local moments outside the FMP into account.  The predicted 
size is therefore too small in the critical regime and the energy is 
overestimated.  

\section{Band states} 
\label{varcalcband} 
We check whether the FMP we discovered in the previous section is bound by 
comparing its energy with the position of the band edge. We determine the 
position of the band edge as the lowest energy of a  typical ``band 
states''.  The energy of a band state is the  difference between the energy 
of the system with an electron present in that state and the energy of the 
system in the absence of that electron.  Again we use a variational 
approach with a trial wave function similar to 
Eqn.~\ref{polaronNsitewavefunction}, with the caveat that local moments are 
not aligned.  Only the kinetic and the $s$-$d$ term of the Hamiltonian in 
Eqn.~\ref{petersH} contribute to the energy, the Heisenberg term is the 
same regardless of the presence of the carrier in a band state.  The 
kinetic energy of a localised state is determined as before.  The magnetic 
energy is estimated as follows.  There is no net magnetisation in the 
paramagnet, so the $s$-$d$ term in the Hamiltonian vanishes on average.  In 
a finite region however, there are statistical fluctuations that make it 
deviate from this average value, so that there is a non-zero magnetisation 
in this region, 
$\vec{m}_{\mathrm{fluct}}= \sum_i \rho_{\vec{r}_i} \vec{S}_{\vec{r}_i}$.  
The carrier's spin is quantised along the direction of 
$\vec{m}_{\mathrm{fluct}}$, pointing ``up''.  The average value of the $s$-$d$
term in the Hamiltonian is then
\begin{equation}
E_{s-d} = -\frac{J'}{2}\left\langle\sqrt{
                                       \sum_{ij} \rho_{\vec{r}_i}
                                                 \rho_{\vec{r}_j}
                                                 \vec{S}_{\vec{r}_i}\cdot      
                                                 \vec{S}_{\vec{r}_j}          
                                      }
                     \right\rangle.
\label{band sd energy}
\end{equation}
The above sum is split up in one where $i=j$ and one where $i\ne j$.  We 
then take the thermal average of the Taylor expansion of the square root 
about the $i=j$ term.  The sum that contains term with $i\ne j$ is at least 
of order $\beta J$ since it contains correlations between different local 
moments.  The $s$-$d$ energy of the band state is then given by
\begin{equation}
E_{s-d} = -\frac{J'S}{2}\sqrt{ \sum_i \rho_{\vec{r}_i}^2 }\,
[1+{\mathcal{O}}(\beta J)].
\label{band es energy 2}
\end{equation}
The energy of the band state is minimised numerically with respect to the 
size.  We obtain a size of the order of 10$^6$ local moments that decreases 
weakly with increasing $J'$.  We find that the $s$-$d$ energy is completely 
negligible and that the band state is so large that its energy is very 
close to $-6|t|$.  We expand the band state energy to second order in $1/l$ 
for large $l$ and obtain:
\begin{equation}
E_{\mathrm{Band}} = -6|t|(1-\frac{1}{l^2})-\frac{J'S}{2} \sqrt{\frac{1}{(\sqrt{2}l)^3}},
\label{largebandstateenergy}
\end{equation}
where $l$ is the spatial extent of the band state.  The polaron level lies 
therefore below the band edge and the FMP is well bound.

\section{Crossing rate with large Band States}
\label{sec:large band states}
We use the same method to calculate the crossing rate for large band 
states.  However, instead of two spins fluctuating into nearly exact 
alignment many spins need to collectively fluctuate into a more aligned 
configuration.  The time between two crossings is again estimated from the 
time it takes for the RMS deviation of the band state energy to become as 
large as the energy difference between the band level and the polaron level:
\begin{equation}
\left\langle \left\{ E_N[\tau_{\mathrm{X}}(N)] - E_{N}(0) \right\}^2\right\rangle = 
\left(E_{\mathrm{P}} - E_{N}\right)^2.
\label{tauXNdef}
\end{equation}
Here $\tau_{\mathrm{X}}(N)$ is the average time between crossings of the 
polaron level and a particular band state of $N$ local moments with energy  
$E_N$.  In Appendix \ref{varcalcband} we calculate energy of the band state 
using a variational Ansatz with an accurate Gaussian trial wave function to 
show that the FMP is bound.  Here, however, we determine the crossing 
rate and we will see that a simple model is sufficient.  We assume that the 
carrier is localised uniformly without inducing extra alignment.  Its spin 
quantised along the direction of the sum of all the local moments in the 
band state, $\vec{m}$, pointing ``up''.  The kinetic energy of the band 
state is approximately $-6|t|(1-\pi^2/N^{2/3})$ and its exchange energy 
vanishes on average.  The kinetic energy is constant in time, so it cancels 
out in the LHS of Eqn.~\ref{tauXNdef}, leaving only the RMS deviation of 
the exchange energy.

Now we use the expression for the 2-site polaron of Eqn. \ref{2siteenergy} 
on the RHS of Eqn. \ref{tauXNdef}.  The band state $s$-$d$ energy can be 
expressed as $\sigma m(t)/N^2$ and hence Eqn.~\ref{tauXNdef} becomes:
\begin{eqnarray}
&& \left[
      \left\langle
                  \{
                    \sum_{i,j,k,l}^{N} 
                     [\vec{S}_{\vec{r}_i}(t) \cdot \vec{S}_{\vec{r}_j}(t)] 
                     [\vec{S}_{\vec{r}_k}(0) \cdot \vec{S}_{\vec{r}_l}(0)]   
                  \}^{\frac{1}{2}}
      \right\rangle
 \right]_{t=0}^{t=\tau_{\mathrm{X}}(N)}
\!\!\!\!\!\!\! =\nonumber\\
&&\!\!\!\!\!\!\!2N^2 \left[ (5 - \frac{\pi^2}{N^{2/3}})\frac{|t|}{J'} -\frac{S}{2}\right]^2.
\label{eqn for tauXN 2}
\end{eqnarray}
The sum on the LHS of this equation contains $N^2$ terms of the form $S^4$ 
and $2N$ terms of the form 
$S^2\sum_{i\ne j}^N \vec{S}_{\vec{r}_i}\cdot\vec{S}_{\vec{r}_j}$ where the 
spin operators are evaluated at equal times.  There are also terms where 
$i\ne j$ and $k\ne l$.  The square root is expanded about the term of order 
$N$: 
\begin{eqnarray}
\{
 \sum_{i,j,k,l}^{N} 
 [\vec{S}_{\vec{r}_i}(t) \cdot \vec{S}_{\vec{r}_j}(t)] 
 [\vec{S}_{\vec{r}_k}(0) \cdot \vec{S}_{\vec{r}_l}(0)]   
\}^{\frac{1}{2}}
=  \nonumber \\
+\, NS(S+1)+ \sum_{i\ne j}^N\vec{S}_{\vec{r}_k}\cdot\vec{S}_{\vec{r}_l}
 +\nonumber \\
+\sum_{i\ne j,k\ne l}^N
    \!\!\!\!\frac{[\vec{S}_{\vec{r}_i}(t)\cdot\vec{S}_{\vec{r}_j}(t)] 
                 [\vec{S}_{\vec{r}_k}(0)\cdot\vec{S}_{\vec{r}_l}(0)]}
                {2NS(S+1)} + \cdots .
\label{sqrtexpansiontauXN}
\end{eqnarray}
Thermal averaging both sides of this equation results in many two- and 
four-point spin correlators.  We only retain the correlators of lowest 
(quadratic) order in the small parameter $\beta J$, thereby considering 
only nearest neighbour interactions.  Moreover, the first and second terms 
on the RHS of Eqn. \ref{sqrtexpansiontauXN} cancel in Eqn. \ref{eqn for 
tauXN 2} since they do not depend on time.  Hence the condition for a level 
crossing reduces to 
\begin{eqnarray}
   \left\langle(\vec{S}_0 \cdot \vec{S}_1)^2 \!\!\!  -
        \{\vec{S}_0[\tau_{\mathrm{X}}(N)]\cdot\vec{S}_1[\tau_{\mathrm{X}}(N)]\}
               [\vec{S}_0 \cdot \vec{S}_1]
   \right\rangle 
=\nonumber \\
\!\!\!\!\frac{2S(S+1)N^2}{3} \left[ (5 - \frac{\pi^2}{N^{2/3}})\frac{|t|}{J'} -\frac{S}{2}\right]^2.
\label{eqn for tauXN 3}
\end{eqnarray} 
The four-point correlator is treated as in Eqn.~\ref{def of f} and we 
introduce the numerical constant $A$ as in Eqn.~\ref{tauX2}:
\begin{eqnarray}
f(\tau_{\mathrm{X}}(N)) 
&=&A\,
N^2 \frac{2S(S+1)}{3\langle\vec{S}_0\cdot\vec{S}_1\rangle^2}\times\nonumber \\ 
&\times & \left[(5-\frac{\pi^2}{N^{2/3}})\frac{|t|}{ J'}-\frac{S}{2}\right]^2.
\label{eqn for tauXN 5}
\end{eqnarray}

$f(t)$ could be expanded about $t=\tau_{\mathrm{S}}$ to obtain an explicit 
expression for $\tau_{\mathrm{X}}(N)$.  However, it is clear that large 
band states rarely cross the polaron level.  $f(t)$ on the LHS is bounded 
from above by 1 and the RHS is proportional to $N^2$, and so only small 
band states satisfy Eqn. \ref{eqn for tauXN 5}.  Since $\tau_N$ grows with 
$N$, the size of the largest crossing band state follows from Eqn.~\ref{eqn 
for tauXN 3} in the limit $\tau_{\mathrm{X}}(N) \to \infty$.  In this limit 
$f(t)=1$ so that the largest crossing band state needs 
$N\sim \langle\vec{S}_0\cdot\vec{S}_1\rangle^2 \ll 1$.    
This result shows that large band states do not cross the polaron level 
according to Gaussian statistics.  The physical reason is that the 
contribution to the $s$-$d$ energy of a single pair of local moments is 
weighted by the electron density.  For large states this density is low, so 
that alignment of a single pair of local moments does not lower the energy 
significantly.  Thus, a crossing requires all the local moments to fluctuate 
into nearly perfect alignment.  Of course, large clusters of 
ferromagnetically aligned local moments do exist, and do cross the polaron 
level, but they lie in the far tail of the band and are far more rare than 
a Gaussian approximation to the density of states would predict. These 
fluctuations are thus negligible.
We have therefore shown that our expressions for $\tau_\mathrm{X}$ in Eqn 
(\ref{tauX3a}) and (\ref{tauX3b}) give the correct hopping rate for the small FMP.

\end{document}